\documentclass{iopart}
\usepackage{cite}
\usepackage{graphicx}
\usepackage{latexsym}

\newlength{\figurewidth}
\begin{document}

\paper{Synchronization of random walks with reflecting boundaries}
\author{Andreas Ruttor, Georg Reents and Wolfgang Kinzel}
\address{Institut f\"ur Theoretische Physik, Universit\"at W\"urzburg,
  Am Hubland, 97074 W\"urzburg, Germany}
\eads{\mailto{andreas.ruttor@physik.uni-wuerzburg.de},
  \mailto{georg.reents@physik.uni-wuerzburg.de} and
  \mailto{wolfgang.kinzel@physik.uni-wuerzburg.de}}

\begin{abstract}
  Reflecting boundary conditions cause two one-dimensional random
  walks to synchronize if a common direction is chosen in each step.
  The mean synchronization time and its standard deviation are
  calculated analytically. Both quantities are found to increase
  proportional to the square of the system size. Additionally, the
  probability of synchronization in a given step is analyzed, which
  converges to a geometric distribution for long synchronization
  times. From this asymptotic behavior the number of steps required to
  synchronize an ensemble of independent random walk pairs is deduced.
  Here the synchronization time increases with the logarithm of the
  ensemble size. The results of this model are compared to those
  observed in neural synchronization.
\end{abstract}


\section{Introduction}

Synchronization of neural networks by mutual learning
\cite{Metzler:2000:INN, Kinzel:2000:DIN} has recently been applied to
cryptography \cite{Kanter:2002:SEI}. Two neural networks which are
trained by their mutual output bits can synchronize to a state with
common time dependent couplings \cite{Kinzel:2002:INN}. It has been
shown how this phenomenon can be used to generate a secret key over a
public channel \cite{Mislovaty:2002:SKE, Kinzel:2003:DGI}.

The networks used for neural cryptography are tree parity machines
\cite{Rosen-Zvi:2002:MLT} consisting of $K$ hidden units, which are
discrete perceptrons with independent receptive fields made up of $N$
binary input neurons. The influence of each input value on the output
of the neural network is determined by the corresponding weight, which
is an integer number in the range between $-L$ and $+L$. Since the
synaptic depth of the neural networks is limited by the parameter $L$,
there are only $m=2L+1$ possible values for each weight.

In each step a set of common inputs is chosen randomly for all tree
parity machines involved in the process of synchronization. Then the
weights are updated according to a suitable learning rule
\cite{Ruttor:2004:NCF}. Thereby the change of each weight can be
either $+1$, $0$ or $-1$ depending on the corresponding input value
and the calculated output of the neural networks. If an update is not
possible because of the limited synaptic depth, then the affected
weight is left unchanged in this step. That is why the process of
synchronization is identical to an ensemble of random walks with
reflecting boundaries, driven by pairwise identical random signals and
controlled by mutual output bits \cite{Kinzel:2002:INN,
  Kinzel:2003:DGI}.

Therefore, to better understand neural cryptography, it is the aim of
this paper to investigate the process of synchronization of random
walks in some detail. We neglect the effect of the control bits, which
is essential for cryptography \cite{Klimov:2002:ANC}, but we derive
exact solutions for the synchronization times. In addition, since
synchronization is an active field of research (for an overview see
\cite{Pikovsky:2001:S}), synchronization of random walks may be of
interest in other contexts, too.

\begin{figure}
  \centering
  \includegraphics[width=\figurewidth]{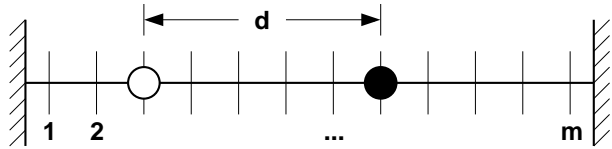}
  \caption{Random walks with reflecting boundaries.}
  \label{fig:walk}
\end{figure}

In this paper we analyze the model shown in \fref{fig:walk}. Two
random walks can move on a one-dimensional line with $m$ sites. In
each step a direction, either left or right, is chosen randomly. Then
both random walkers move in this direction. If one random walk hits
the boundary, it is reflected, i.e. the random walk remains at its
site on the left or on the right. As the other random walker is not
affected, the distance $d$ between both decreases by $1$ at each
reflection.  Otherwise $d$ remains constant.

The important quantity in this model is the synchronization time $T$,
which is defined as the number of steps until $d$ reaches zero. In
\sref{sec:average} we calculate the mean value and the standard
deviation of this random variable analytically. Because large
fluctuations of the synchronization time are observed in simulations,
the average value of $T$ does not describe the behavior of the model
sufficiently. Therefore, we estimate the probability distribution of
the synchronization time in \sref{sec:dist}. Using this result we are
able to calculate the number of steps $T_N$, after which an ensemble
of $N$ random walk pairs reaches a synchronized state. This quantity
is equal to the maximum of $T$ observed in $N$ independent samples.
Therefore, we use methods of extreme order statistics to calculate the
distribution of $T_N$, which is shown in \sref{sec:max}.

\section{Average synchronization time}
\label{sec:average}

In order to calculate the mean value $\langle T \rangle$ of the
synchronization time, the syn\-chro\-ni\-za\-tion process is divided
into independent parts, each of them with constant distance $d$. We
first compute the average number $\langle S_{d,z} \rangle$ of steps
until a reflection occurs, which decreases $d$ by $1$. Of course, this
value depends on the initial position of the two random walkers. We
use the distance $d$ between both walkers and the position $z$ of the
left walker to describe these initial conditions.

If the first move is to the right, a reflection only occurs in the
case of $z=m-d$. Otherwise the synchronization process continues as if
the initial position had been $z+1$. Under this condition, the average
number of steps with distance $d$ is given by $\langle S_{d,z+1} + 1
\rangle$. Similarly, if the two random walkers move to the left in the
first step, this quantity is equal to $\langle S_{d,z-1} + 1 \rangle$.
Averaging over both possibilities we obtain the following difference
equation:
\begin{equation}
  \label{eq:difference}
  \langle S_{d,z} \rangle = \frac{1}{2} \langle S_{d,z-1} \rangle +
  \frac{1}{2} \langle S_{d,z+1} \rangle + 1 \, .
\end{equation}

Reflections are only possible if the current position $z$ is either
$1$ or $m-d$. In both situations $d$ changes with probability
$\frac{1}{2}$ in the next step. In order to take this into account we
have to use
\begin{equation}
  \label{eq:boundary}
   S_{d,0}=0 \quad \mbox{and} \quad S_{d,m-d+1}=0
\end{equation}
as boundary conditions.

\Eref{eq:difference} is identical to the classical ruin problem
\cite{Feller:1968:IPT}. Its solution is:
\begin{equation}
  \label{eq:solution}
  \langle S_{d,z} \rangle = (m - d + 1) z - z^2 \, .
\end{equation}

In a similar manner we obtain a difference equation for $\langle
S_{d,z}^2 \rangle$, which is used later to calculate the standard
deviation of the synchronization time:
\begin{equation}
  \label{eq:diff2}
  \langle S_{d,z}^2 \rangle =  \frac{1}{2} \langle (S_{d,z-1} + 1)^2
  \rangle + \frac{1}{2} \langle (S_{d,z+1} + 1)^2 \rangle \, .
\end{equation}
From (\ref{eq:solution}) and (\ref{eq:diff2}) we obtain the following
relation for the variance of $S_{d,z}$:
\begin{equation}
    \fl \langle S_{d,z}^2 \rangle - \langle S_{d,z} \rangle^2 =
    \frac{\langle S_{d,z-1}^2 \rangle - \langle S_{d,z-1}
      \rangle^2}{2} + \frac{\langle S_{d,z+1}^2 \rangle - \langle
      S_{d,z+1} \rangle^2}{2} + \left( m - d + 1 - 2 z \right)^2 .
\end{equation}
Applying a Z-transformation yields the solution
\begin{equation}
  \langle S_{d,z}^2 \rangle - \langle S_{d,z} \rangle^2 = \frac{(m - d
    + 1 - z)^2 + z^2 - 2}{3} \, \langle S_{d,z} \rangle \, .
\end{equation}

With these results we can calculate the average value of the
synchronization time $T_{d,z}$ for two random walks starting at
distance $d$ and position $z$. The first reflection occurs after
$S_{d,z}$ steps. Then one of the random walkers is located at the
boundary. As our model is symmetric, both possibilities $z=1$ or
$z=m-d$ are equal. Hence the second reflection takes place after
$S_{d,z} + S_{d-1,1}$ steps. So the total synchronization time is
given by
\begin{equation}
  \label{eq:time}
  T_{d,z} = S_{d,z} + \sum_{j=1}^{d-1} S_{j,1} \, .
\end{equation}
Using \eref{eq:solution} we obtain
\begin{equation}
  \langle T_{d,z} \rangle = (m - d + 1) z - z^2 + \, \frac{1}{2} (d -
  1)(2 m - d)
\end{equation}
for the expectation value of this random variable. The variance of
$T_{d,z}$ can be calculated in a similar manner, because the parts of
the synchronization process are mutually independent.

Finally, we have to average over all possible initial conditions in
order to calculate the mean value and the standard deviation of the
synchronization time $T$ for randomly chosen starting positions of the
two random walkers:
\begin{eqnarray}
  \label{eq:t1}
  \fl \langle T \rangle &= \frac{2}{m^2} \sum_{d=1}^{m-1}
  \sum_{z=1}^{m-d} \langle T_{d,z} \rangle &= \frac{(m-1)^2}{3} +
  \frac{m-1}{3m} \, , \\
  \label{eq:t2}
  \fl \langle T^2 \rangle &= \frac{2}{m^2} \sum_{d=1}^{m-1}
  \sum_{z=1}^{m-d} \langle T_{d,z}^2 \rangle &= \frac{17 m^5 - 51 m^4
    + 65 m^3 - 45 m^2 + 8 m + 6}{90 m} \, .
\end{eqnarray}

\begin{figure}
  \centering
  \includegraphics[width=\figurewidth]{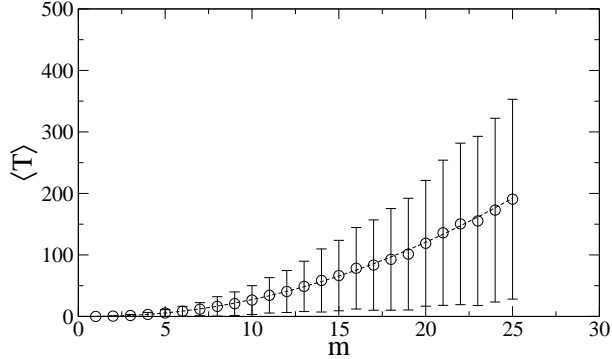}
  \caption{Synchronization time of two random walks as a function of
    the system size $m$. Error bars denote the standard deviation
    observed in 1000 simulations. The analytical solution from
    \eref{eq:t1} is plotted as dashed curve.}
  \label{fig:time}
\end{figure}

So the average number of steps required to reach a synchronized state,
which is shown in \fref{fig:time}, increases nearly proportional to
$m^2$. In particular for large system sizes $m$, we obtain the
asymptotic relation
\begin{equation}
  \label{eq:mean}
  \langle T \rangle \sim \frac{1}{3} m^2 \sim \frac{4}{3} L^2 \, .
\end{equation}
This result is consistent with the scaling behavior $\langle T \rangle
\propto L^2$ observed in neural synchronization
\cite{Mislovaty:2002:SKE}.

Numerical simulations, both for random walks and neural networks, show
large fluctuations of the synchronization time. The reason for this
observation is that not only the mean value, but also the standard
deviation
\begin{equation}
  \sigma_T = \sqrt{\frac{7 m^6 - 11 m^5 - 15 m^4 + 55 m^3 - 72 m^2 +
      46 m - 10}{90 m^2}}
\end{equation}
of the synchronization time increases with the extension $m$ of the
random walks. Here we find that $\sigma_T$ is asymptotically
proportional to $\langle T \rangle$:
\begin{equation}
  \sigma_T \sim \sqrt{\frac{7}{10}} \, \langle T \rangle \, .
\end{equation}
Therefore, the relative fluctuations $\sigma_T / \langle T \rangle$
are nearly independent of $m$ and not negligible.

\section{Probability distribution}
\label{sec:dist}

The synchronization of two random walks is a Markov chain, because the
next step only depends on the current state of the system. We can,
therefore, derive an equation for the distribution of the total
synchronization time $T$.

For this purpose we introduce $\frac{1}{2} m (m + 1)$ variables
$p_{d,z}(t)$, which are defined as the probability to find the two
random walkers at the positions $z$ and $z+d$ in time step $t$. In the
initial ensemble the positions are chosen randomly, hence all
variables are set to
\begin{equation}
  p_{d,z}(0) = \frac{2 - \delta_{d,0}}{m^2} \, .
\end{equation}
The development of the probabilities in each step is given by the
following equations for $0 \leq d < m-1$ and $1 < z < m-d$:
\begin{eqnarray}
  \label{eq:dyn1}
  p_{d,z}(t+1) &=&
  \frac{1}{2} \, p_{d,z-1}(t) + \frac{1}{2} \, p_{d,z+1}(t) \, , \\
  \label{eq:dyn2}
  p_{d,1}(t+1) &=&
  \frac{1}{2} \, p_{d,2}(t) + \frac{1}{2} \, p_{d+1,1}(t) +
  \frac{1}{2} \, \delta_{d,0} \, p_{0,1} \, , \\
  \label{eq:dyn3}
  p_{d,m-d}(t+1) &=&
  \frac{1}{2} \, p_{d,m-d-1}(t) + \frac{1}{2} \, p_{d+1,m-d-1}(t) +
  \frac{1}{2} \, \delta_{d,0} \, p_{0,m} \, , \\
  \label{eq:dyn4}
  p_{m-1,1}(t+1) &=& 0 \, .
\end{eqnarray}

The probability $\mathrm{P}(T=t)$, that the two random walks
synchronize in the time step $t>0$, is given by
\begin{equation}
  \mathrm{P}(T=t) = \sum_{z=1}^{m} \left( p_{0,z}(t) - p_{0,z}(t-1)
  \right) \, .
\end{equation}
Using (\ref{eq:dyn1})--(\ref{eq:dyn4}) we deduce the expression
\begin{equation}
  \mathrm{P}(T=t) = \frac{1}{2} \, p_{1,1}(t-1) + \frac{1}{2} \,
  p_{1,m-1}(t-1) \, .
\end{equation}
These equations can easily be iterated numerically.

\begin{figure}
  \centering
  \includegraphics[width=\figurewidth]{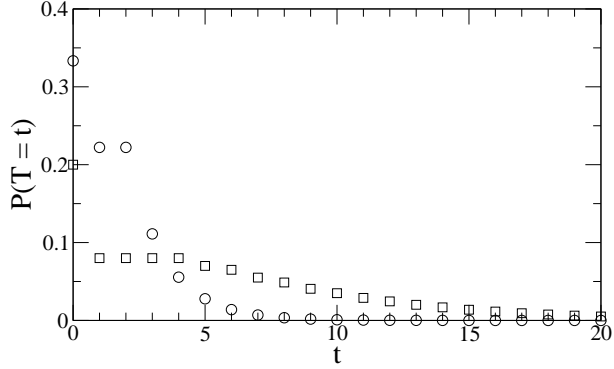}
  \caption{Probability of synchronization $\mathrm{P}(T=t)$ as a
    function of the number of steps. The symbols show the result of
    the numerical calculation for $m=3$ ($\bigcirc$) and $m=5$
    ($\Box$).}
  \label{fig:rwdist}
\end{figure}

\Fref{fig:rwdist} shows the results of these calculations for $m=3$
and $m=5$. Because of the randomly chosen initial conditions there is
a probability of $\mathrm{P}(T=0) = 1 / m$ that the two random walkers
even start synchronized. Furthermore, one notices that
$\mathrm{P}(T=t)$ is constant and equal to $2 / m^2$ in the range $0 <
t < m$.

In principle, it is also possible to calculate the probability
distribution $\mathrm{P}(T=t)$ analytically. We use this method to
estimate an approximation of $\mathrm{P}(T=t)$, which is
asymptotically exact for very long synchronization times $t \gg m$.
For that purpose we start with a result known from the solution of the
classical ruin problem \cite{Feller:1968:IPT}: The probability that a
fair game ends with the ruin of one player in time step $t$ is given
by
\begin{equation}
  \label{eq:u}
  \fl u(t) = \frac{1}{a} \sum_{k=1}^{a-1} \sin \left( \frac{k \pi
      z}{a} \right) \left[ \sin \left( \frac{k \pi}{a} \right) + \sin
    \left( k \pi - \frac{k \pi}{a} \right) \right] \left[ \cos \left(
      \frac{k \pi}{a} \right) \right]^{t-1} \, .
\end{equation}
In our model $a - 1 = m - d$ denotes the number of possible positions
for two random walkers with distance $d$. And $u(t)$ is the
probability distribution of the random variable $S_{d,z}$, which we
introduced in \sref{sec:average}. As before, $z$ is the initial
position of the left random walker.

According to \eref{eq:time} the synchronization time $T_{d,z}$ for
fixed initial conditions is the sum over $S_{i,j}$ for each distance
$i$ from $d$ to $1$. Therefore, its probability distribution
$\mathrm{P}(T_{d,z}=t)$ is a convolution of $d$ functions $u(t)$
defined in \eref{eq:u}. The convolution of two different geometric
sequences $b_n = b^n$ and $c_n = c^n$ is itself a linear combination
of these sequences:
\begin{equation}
  \label{eq:gc}
  b_n \ast c_n = \sum_{j=1}^{n-1} b^j c^{n-j} = \frac{c}{b-c} \, b_n +
  \frac{b}{c-b} \, c_n \, .
\end{equation}
Therefore, $\mathrm{P}(T_{d,z}=t)$ can be written as a sum over
geometric sequences, too:
\begin{equation}
  \mathrm{P}(T_{d,z}=t) = \sum_{a=m - d + 1}^{m} \sum_{k=1}^{a-1}
  q^{d,z}_{a,k} \left[ \cos \left( \frac{k \pi}{a} \right)
  \right]^{t-1} \, .
\end{equation}

In order to obtain $\mathrm{P}(T=t)$ for random initial conditions, we
have to average over all possible starting positions of the random
walkers:
\begin{equation}
  \mathrm{P}(T=t) = \frac{2}{m^2} \sum_{d=1}^{m-1} \sum_{z=1}^{m-d}
  \mathrm{P}(T_{d,z}=t) \, .
\end{equation}
So we can even write
\begin{equation}
  \mathrm{P}(T=t) = \sum_{a=2}^{m} \sum_{k=1}^{a-1} q_{a,k} \left[
    \cos \left( \frac{k \pi}{a} \right) \right]^{t-1}
\end{equation}
as a sum over many geometric sequences.

\begin{figure}
  \centering
  \includegraphics[width=\figurewidth]{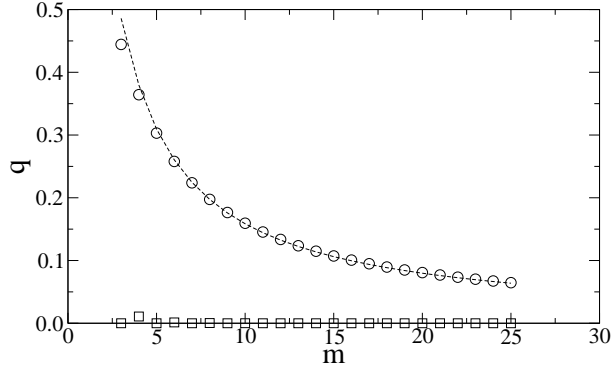}
  \caption{Value of the coefficients $q_{m,1}$ ($\bigcirc$) and
    $q_{m,m-1}$ ($\Box$) as a function of $m$. The parameter $q_{m,1}$
    can be approximately calculated using the function $q_{m,1}
    \approx 0.324 m [1 - \cos (\pi / m)]$, which is shown as dashed
    curve.}
  \label{fig:coeff}
\end{figure}

For long times only the terms with the largest absolute value of the
coefficient $\cos (k \pi / a)$ are relevant, because the others
decline exponentially faster and can be neglected for $t \rightarrow
\infty$. Hence the asymptotic behavior of the probability distribution
is given by
\begin{equation}
  \label{eq:dist}
  \mathrm{P}(T=t) \sim [ q_{m,1} + (-1)^{t-1} q_{m,m-1} ]
  \left[ \cos \left( \frac{\pi}{m} \right) \right]^{t-1} \, .
\end{equation}
The two coefficients $q_{m,1}$ and $q_{m,m-1}$ in this equation can be
calculated using \eref{eq:gc}. This leads to the following result:
\begin{eqnarray}
  \label{eq:q1}
  \fl q_{m,1} = \frac{\sin^2 (\pi / m)}{m^2 m!} \sum_{d=1}^{m-1}
  \frac{2^{d+1} (m - d)!}{1 - \delta_{d,1} \cos (\pi / m)} \nonumber \\
  \times \prod_{a=m-d+1}^{m-1} \sum_{k=1}^{a-1} \frac{\sin^2 (k \pi /
    2)}{\cos (\pi / m) - \cos (k \pi / a)} \, \frac{\sin^2 (k \pi /
    a)}{1 - \delta_{a,m-d+1} \cos (k \pi / a)} \, , \\
  \label{eq:q2}
  \fl q_{m,m-1} = \frac{\sin^2 (\pi / m) \cos^2 (m \pi / 2)}{m^2 m!}
  \sum_{d=1}^{m-1} (-1)^{d-1} \frac{2^{d+1} (m - d)!}{1 + \delta_{d,1}
    \cos (\pi / m)} \nonumber \\
  \times \prod_{a=m-d+1}^{m-1} \sum_{k=1}^{a-1} \frac{\sin^2 (k \pi /
    2)}{\cos (\pi / m) + \cos (k \pi / a)} \, \frac{\sin^2 (k \pi /
    a)}{1 - \delta_{a,m-d+1} \cos (k \pi / a)} \, ,
\end{eqnarray}
which is also shown in \fref{fig:coeff}.

For odd values of $m$, the coefficient $q_{m,m-1} = 0$. In this case
$\mathrm{P}(T=t)$ as\-ymp\-toti\-cal\-ly converges to a geometric
probability distribution for long synchronization times:
\begin{equation}
  \label{eq:asymp}
  \mathrm{P}(T=t) \sim q_{m,1} \left[ \cos \left( \frac{\pi}{m}
    \right) \right]^{t-1} \, .
\end{equation}

\begin{figure}
  \centering
  \includegraphics[width=\figurewidth]{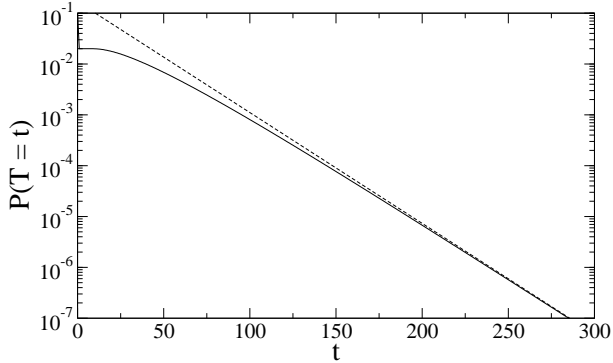}
  \caption{Probability distribution $\mathrm{P}(T=t)$ of the
    synchronization time for $m=10$. The numerical result is plotted
    as full curve. The dashed line denotes the asymptotic function
    defined in \eref{eq:asymp}.}
  \label{fig:asymp}
\end{figure}

If $m$ is even, we have to consider oscillations of $\mathrm{P}(t)$
around the function given by \eref{eq:asymp}, because $q_{m,m-1} \not=
0$. But for $m > 8$, the coefficient $q_{m,m-1}$, which determines the
amplitude of these oscillations, is smaller than $10^{-3} q_{m,1}$. As
shown in \fref{fig:asymp}, \eref{eq:asymp} is a good approximation of
the asymptotic behavior for even values of $m$, too.

\section{Extreme order statistics}
\label{sec:max}

In this section we extend our model to ensembles of random walks. We
consider $N$ independent pairs of random walkers driven pairwise by
identical random noise. Each pair is finally synchronized by the
effect of the reflecting boundary conditions. From the results
discussed above we know the expected average synchronization time. But
the important quantity in this case is the number of steps $T_N$ until
all random walks are synchronized. Of course, this random variable is
equal to the maximum value of $T$ observed in $N$ independent samples.

From the distribution function $\mathrm{P}(T \leq t)$ we can easily
deduce the probability distribution of $T_N$:
\begin{equation}
  \label{eq:pmax}
  \mathrm{P}(T_N \leq t) = \mathrm{P}(T \leq t)^N \, .
\end{equation}

\begin{figure}
  \centering
  \includegraphics[width=\figurewidth]{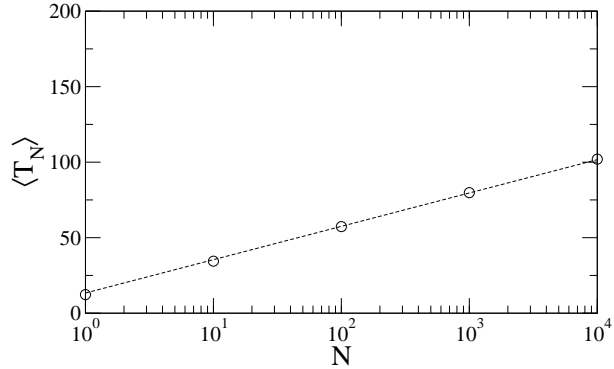}
  \caption{Average synchronization time $\langle T_N \rangle$ as a
    function of $N$ for $m=7$. Results of the numerical calculation
    using \eref{eq:pmax} are represented by circles. The dashed line
    shows the expectation value of $T_N$ calculated in
    \eref{eq:gumexp}.}
  \label{fig:maximum}
\end{figure}

Therefore, we can calculate the average value $\langle T_N \rangle$
using the numerically computed distribution. The result, which is
shown in \fref{fig:maximum}, indicates that $\langle T_N \rangle$
increases proportional to $\ln N$:
\begin{equation}
  \label{eq:maximum}
  \langle T_N \rangle - \langle T \rangle \propto \ln N \, .
\end{equation}
Similar behavior has been observed in neural synchronization
\cite{Kinzel:2002:INN, Kinzel:2003:DGI}.

For large $N$ only the asymptotic behavior of $\mathrm{P}(T \leq t)$
is relevant for the distribution of $T_N$. The exponential decay of
$\mathrm{P}(T=t)$ in \eref{eq:dist} yields a Gumbel distribution for
$\mathrm{P}(T_N \leq t)$ \cite{Galambos:1940:ATE},
\begin{equation}
  \label{eq:gum}
  G(x) = \exp \left( -\rme^\frac{\alpha - x}{\beta} \right) \, ,
\end{equation}
for $N \gg m$ with the parameters
\begin{equation}
  \label{eq:gumpar}
  \alpha = \beta \ln \frac{N q_{m,1}}{1 - \cos (\pi / m)}
  \quad \mbox{and} \quad
  \beta = - \frac{1}{\ln \cos (\pi / m)} \, .
\end{equation}
Substituting \eref{eq:gumpar} into \eref{eq:gum} we get
\begin{equation}
  \mathrm{P}(T_N \leq t) = \exp \left( - \frac{N q_{m,1} \cos^t
      (\pi / m)}{1 - \cos (\pi / m)} \right)
\end{equation}
as the distribution function for the total synchronization time of $N$
pairs of random walks ($N \gg m$). The expectation value of this
probability distribution is given by \cite{Galambos:1940:ATE}
\begin{equation}
  \label{eq:gumexp}
  \langle T_N \rangle = \alpha + \beta \gamma = - \frac{1}{\ln \cos
    (\pi / m)} \left( \gamma + \ln N + \ln \frac{q_{m,1}}{1 -
      \cos (\pi / m)} \right) .
\end{equation}
In this equation, $\gamma$ denotes the Euler-Mascheroni constant. For
$N \gg m \gg 1$, we obtain the relation
\begin{equation}
  \label{eq:approx}
  \langle T_N \rangle \sim \frac{2}{\pi^2} \, m^2 \left( \gamma + \ln
    N + \ln \frac{2 m^2 \, q_{m,1}}{\pi^2} \right) \, ,
\end{equation}
which shows that $\langle T_N \rangle$ increases asymptotically
proportional to $m^2 \ln N$. Using the approximation for $q_{m,1}$
shown in \fref{fig:coeff}, we find $\langle T_N \rangle \approx (2 /
\pi^2) m^2 \left( \ln N + \ln (0.577 m) \right)$.

\begin{figure}
  \centering
  \includegraphics[width=\figurewidth]{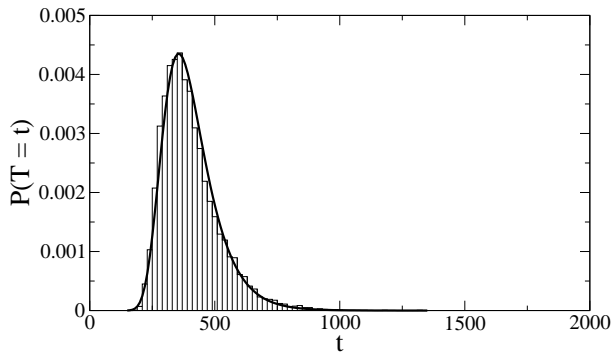}
  \caption{Probability distribution of the synchronization time for
    two tree parity machines with the parameters $K=3$, $L=3$ and
    $N=1000$. The histogram shows the relative frequency of occurrence
    observed in $10\,000$ simulations. The thick curve represents a
    Gumbel distribution for $\alpha=355.8$ and $\beta=84.5$.}
  \label{fig:neurodist}
\end{figure}

Neural cryptography is somewhat more complex than our model with
random walks. Since in this case the movements of the random walks are
controlled by learning rules \cite{Ruttor:2004:NCF}, there are also
steps without changes of the weights. These are not included in our
calculation of the synchronization time. Additionally, repulsive steps
destroying synchronization \cite{Rosen-Zvi:2002:MLT} are possible,
too. Nevertheless, the synchronization time of neural synchronization
scales like $\langle T_N \rangle \propto L^2 \ln N$
\cite{Mislovaty:2002:SKE}, in agreement with \eref{eq:approx}. And its
probability distribution, obtained from numerical simulations and
shown in \fref{fig:neurodist}, is described well by a Gumbel
distribution where the two parameters $\alpha$ and $\beta$ are fitted
to the data.

\section{Conclusion}

We have analyzed the synchronization of two random walks with
reflecting boundary conditions. We calculated the mean value and the
standard deviation of the synchronization time $T$ analytically for
randomly chosen initial positions of the two random walkers. The
average number of steps until synchronization increases with $m^2$,
the square of the system size. Additionally, the standard deviation
$\sigma_T$ also scales with $m^2$, which shows that fluctuations of
$T$ cannot be neglected even if $m$ is large.

The probability $\mathrm{P}(T=t)$ that two random walks synchronize at
$t$ steps has been derived.  For long synchronization times $t \gg m$
the asymptotic behavior of $\mathrm{P}(T=t)$ is given by a geometric
probability distribution with parameter $p = 1 - \cos (\pi / m)$.

We have also studied the number of steps $T_N$ needed to synchronize
$N$ independent pairs of random walks. The average value of this
random variable increases with $\ln N$. And, for large values of $N$,
the probability $\mathrm{P}(T_N \leq t)$ that $t$ steps are sufficient
for synchronization is given by a Gumbel distribution.

Finally, our model is able to reproduce the scaling behavior observed
in neural cryptography. We find that $T_N$ increases nearly
proportional to $L^2 \ln N$. Additionally, even for tree parity
machines, the distribution of the synchronization times of neural
cryptography is described by a Gumbel distribution.

\section*{References}

\bibliography{paper}
\bibliographystyle{iopp}

\end{document}